\documentclass[aps,prl,twocolumn,showpacs,floatfix,a4paper]{revtex4}
\usepackage{graphicx,amsfonts}
\usepackage[intlimits]{amsmath}
\begin{document}
\title{Persistent current in an almost staggered 
Harper model}
\author{A. Vasserman and R. Berkovits}
\affiliation{Department of Physics, Bar-Ilan
University, Ramat-Gan 52900, Israel}
%
%

\begin{abstract}
In this paper we study the persistent current (PC) of a
staggered Harper model, close to the half-filling.
The Harper model, which is a quasi-periodic system, is different 
than other one dimensional systems with uncorrelated disorder
in the fact that it can be in the metallic regime.
Nevertheless, the PC for a wide range of parameters of
the Harper model
does not show typical metallic behavior, although the system
is in the metallic regime. 
This is a result of the nature of the
central band states, which are a hybridization 
of Gaussian states localized in superlattice points. 
When the superlattice is not commensurate
with the system length, the PC 
behaves as in an insulator. Thus even in the metallic regime 
a typical finite Harper model may exhibit a PC expected from
an insulator.
\end{abstract}
%
\maketitle

In the last decade,
quasi-periodic one-dimensional (1D) potential, 
also known as the Aubry-Andre-Harper (for short
Harper) model \cite{harper55,hofstadter76,aubry80,hiramoto92}, 
has garnered much interest. The main reason for 
this interest is that it is the disorder of choice for cold atoms, 
since it may be created by
the superposition of two incommensurate periodic potentials 
\cite{roati08,chabe08,modugno10,modugno09,lahini09,tanzi13,errico14}. 
Contrary to white noise disordered systems which are always localized 
for 1D \cite{gang4}, the Harper model shows a 
1D metal-insulator (Anderson) transition for non-interacting 
particles as function of the strength of the potential
\cite{aubry80,hiramoto92,roati08,chabe08,modugno10,modugno09,lahini09}.
This has been demonstrated experimentally in cold atoms
\cite{roati08,chabe08,modugno10}, as well as for optical 
systems \cite{modugno09,lahini09}.
Additional effort has gone into understanding the influence of 
electron-electron interactions
on this metal-insulator transition \cite{vidal99,vidal01,schuster02},
and on the the many-body localization in these quasiperiodic system
\cite{iyer13,michal14}.
The Harper model also exhibits topological edge state \cite{kraus12,verbin13}, 
and show counterintuitive behavior of the compressibility 
\cite{kraus14,friedman14}.

The current carried by the ground state of
a ring threaded by a magnetic flux is
called the persistent current (PC) \cite{buttiker83}. 
In the presence of elastic scatterers (disorder),
the persistent current is suppressed 
\cite{cheung88,bouchiat89,vonoppen91,altshuler91}.
In the diffusive regime, on the average it is suppressed
by a power law of the circumference of the ring $L$, while
in the localized regime it suppressed exponentially by
$\exp(-L/\zeta)$, where $\zeta$ is the localization length.
Since for non correlated (white noise) disorder, all
states of a 1D system are localized, one
expects that for 1D systems the averaged PC is always suppressed
exponential.

The PC is a very effective way to evaluate the sensitivity
of the system to boundary conditions, i.e., the conductance
of the system \cite{akkermans92,berkovits96}, and therefore
a great way to identify whether your system is metallic or
localized.
Thus, naively, we would expect that the PC for the Harper model
in the metallic regime (i.e., not too strong on-site potentials) will
on the average be only weakly suppressed by
the potential. Here, we will show that the persistent current
of the Harper model can exhibit a rather intricate behavior,
which can skew the simple picture presented above.

In this paper, we study the PC of a Harper model of spinless
fermions on a ring threaded by a magnetic flux. This is a tight-binding 
model in which the on-site potential is spatially modulated 
with an irrational frequency. 
We would focus here on irrational frequencies which their modulus
is close to half. This
corresponds to a fast (two site) modulation with a slow envelope. 
These frequencies exhibit an increase of the compressibility when the
electron-electron interactions are increased, opposite to
the influence of interactions in regular disordered systems 
\cite{kraus14,friedman14}.
Here we will show that for this range of frequencies close to half-filling, 
the PC shows 
a non-monotonous dependence on the systems size, where for most 
values of $L$ the PC is strongly suppressed.

The tight-binding Harper model Hamiltonian for spinless fermions
on a ring threaded by a magnetic flux is:

\begin{eqnarray} \label{hamiltonian}
H &=& 
\displaystyle \sum_{j=1}^{L} \lambda cos(2\pi bj+\phi){\hat c}^{\dagger}_{j}{\hat c}_{j}
-t \displaystyle \sum_{j=1}^{L-1}(e^{i\varphi/\varphi_0}{\hat c}^{\dagger}_{j}{\hat c}_{j+1} + h.c.) 
\\ \nonumber
\end{eqnarray}

where ${\hat c}_{j}$ is the single particle annihilation operator
on site $j$, $t$ is a real hopping amplitude.
The magnetic flux is denoted by $\varphi$, and
$\varphi_0$ is the quantum flux quanta $\varphi_0=hc/e$.
The strength of the on-site potential is controlled by ${\lambda>0}$. 
The on-site potential 
is modulated by a frequency $b$, and $\phi$ is an arbitrary phase factor.
It should be clear that since we are interested in a ring, $\phi$ 
is irrelevant and will be ignored through the rest of this paper.
We will be interested in the metallic regime of the model, i.e $\lambda<2t$.
The irrational frequency may 
to written as $b=\mathbb{Z}+1/2+\epsilon$, 
where $\epsilon$ is irrational. Therefore, 
we can write the on-site potential as:

\begin{eqnarray} \label{on-site}
\cos(2\pi bj) &= \cos(2\pi \mathbb{Z} j+\pi j+2\pi \epsilon j) \\ \nonumber 
&= (-1)^{j}\cos(2\pi\epsilon j) 
\end{eqnarray}

\begin{figure}
\includegraphics[width=8.5cm]{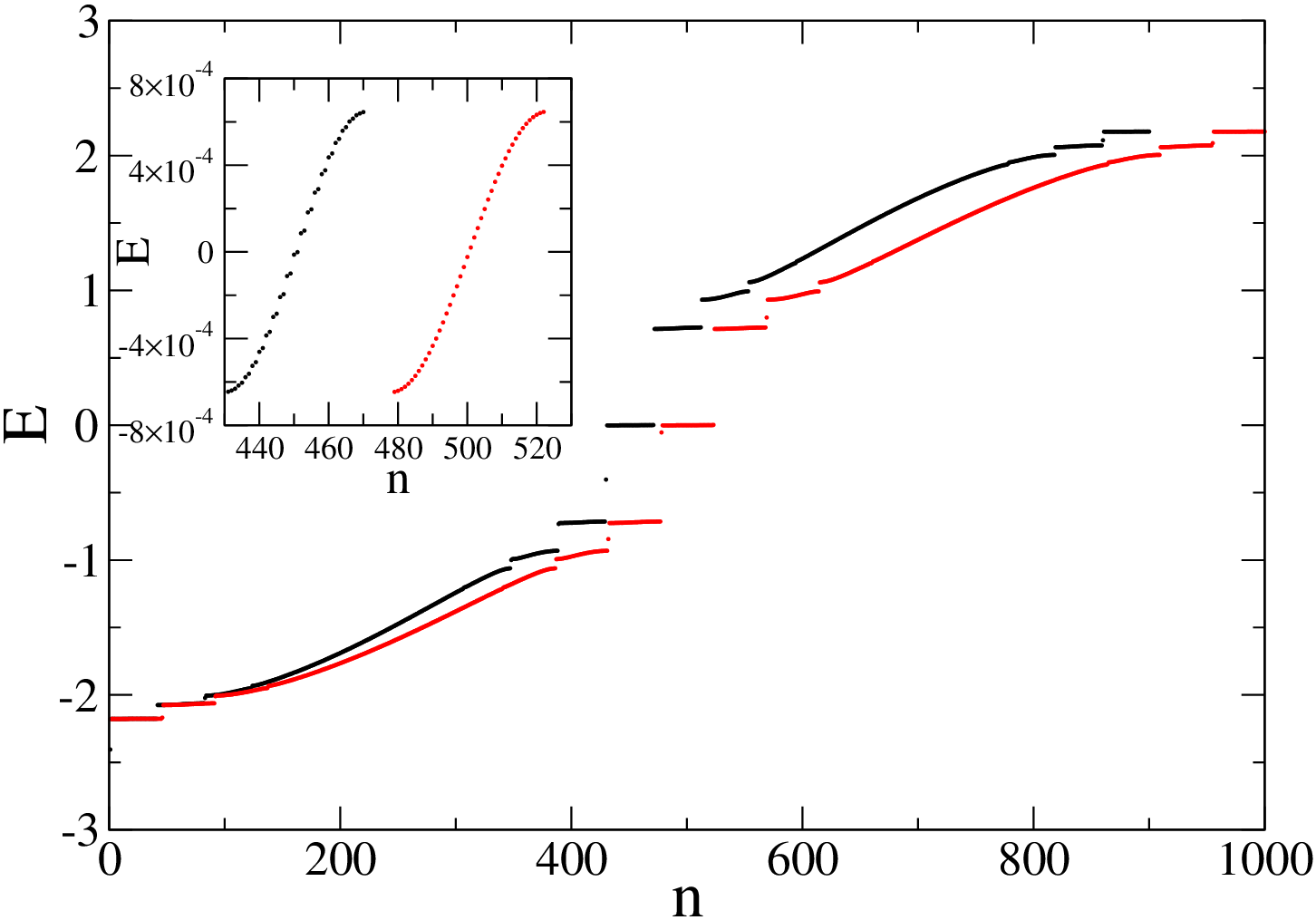}
\caption{\label{fig0}
(Color online)
The energy spectrums of two different length, one with $L=900$ (black symbols)
and the other with $L=1000$ (red symbols), for both length $b=\sqrt{30}$ and 
$\lambda=1$. The superlattie length correspond to
$L_n \sim  50$ for $L=900$ and $L_n \sim  54.55$ for $L=1000$.
The gross features of the spectrum
(except for the edge states appearing in the gaps) do not essentially change
between the commensurate and incommensurate length.
Inset: a zoom into the central bad for both length. Please note the change in
the scale of the y axis. Here there is a clear difference between the 
commensurate and incommensurate systems.
}
\end{figure}

\begin{figure}
\includegraphics[width=8.5cm]{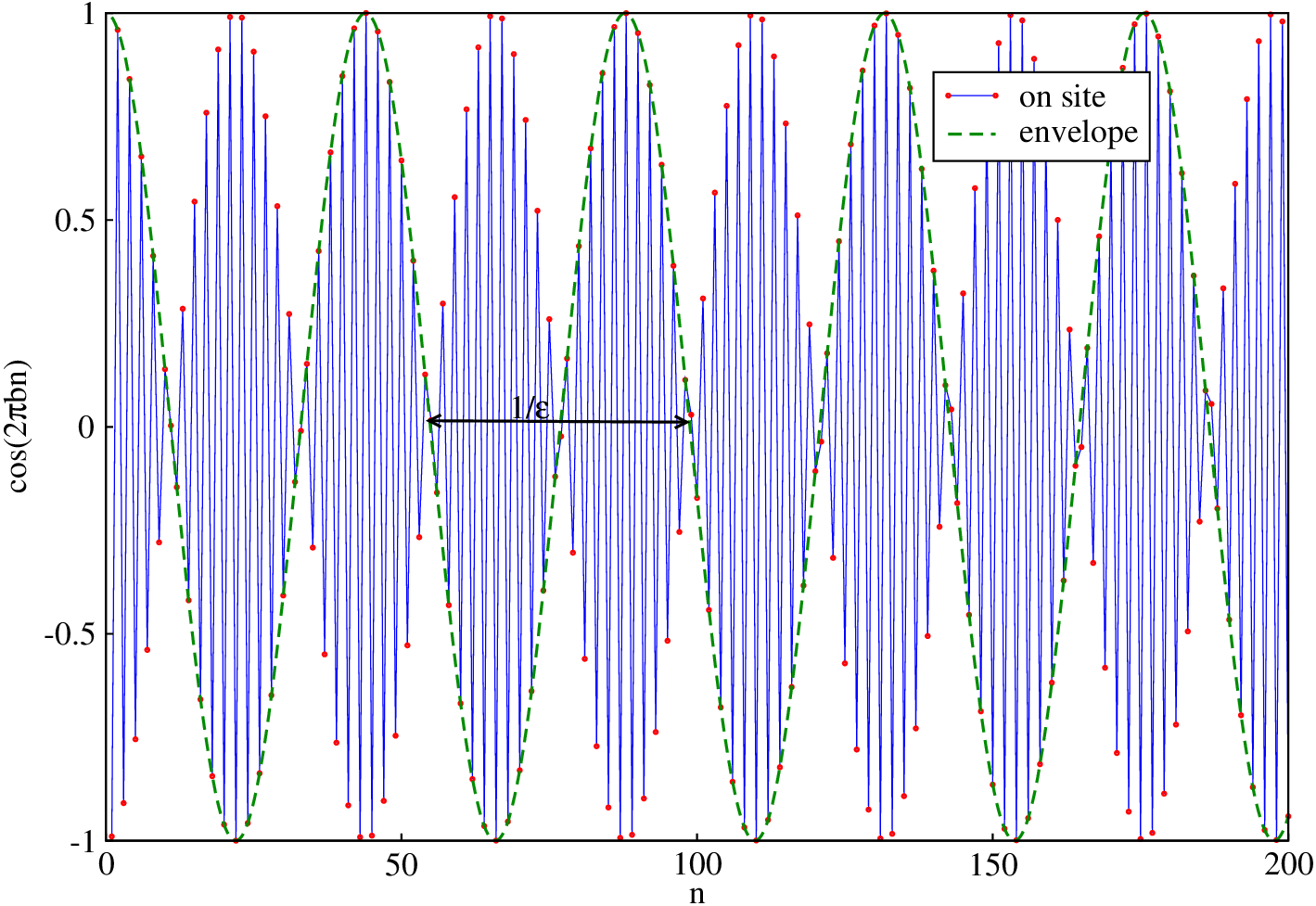}
\vskip 0.2truecm
\caption{\label{fig1}
(Color online)
The on-site potential of the Harper model for $L=200$, with $\lambda=1$ 
and $b=\sqrt{30}$. The envelope corresponds to $\cos(2\pi\epsilon n)$.
The distance between the valleys is $1/2 \epsilon$. The number of
valleys correspond to $L_n = 2|\epsilon|L$ which for the system depicted
in this figure corresponds to $L_n \sim  9.1$, which is a slightly
incommensurate case. 
}
\end{figure}

When $\epsilon \ll 1/2$ the system is called an almost staggered Harper model
for which the fast frequency of the $(-1)^{j}$ term is modulated
by the slow frequency, $\epsilon$, of the $\cos(2\pi\epsilon j)$ term. 
In the almost staggered case the energy 
spectrum of system shows unique features such as a central band that is 
separated from the other bands by a large gap, of order
$\lambda$,
as can be seen in Fig.\ref{fig0}.  
Also the two bands sandwiching the central band show similar features,
i.e., a rather narrow (flat) band and a large gap to the next band.
Changing the length of the sample length from $L=900$ to $L=1000$
does not change its gross features, although some difference in
the energies of the edge states in the gaps are apparent.
As detailed in Ref.  \cite{kraus14}, for $\epsilon \ll 1$, there are 
$L_n = 2|\epsilon|L$ states in the central band, 
corresponding to the number of intersections with zero of the slow 
modulation envelope, which occur at $\cos(2\pi\epsilon j_n)=0$. 
These valleys are shown in Fig.\ref{fig1} for a smaller system
($L=200$) in order not to clutter the figure. 
The frequency of the
envelope is $\epsilon$ and the distance between two consecutive
valleys is half the period, i.e., the distance is $1/2|\epsilon|$. 
For example, for the systems depicted in Fig.\ref{fig0},
$b=\sqrt{30}=5.477226$, and therefore $\epsilon=-0.022774$, resulting in
$L_n \sim  50$ for $L=900$ and $L_n \sim  54.55$ for $L=1000$.
While
for the $L=200$ case shown in Fig. \ref{fig1} $L_n \sim  9.1$. Indeed,
the $9$ valley states are clearly seen, as well as the fact that the
valley positions ate not exactly commensurate.

\begin{figure}
\includegraphics[width=8.5cm]{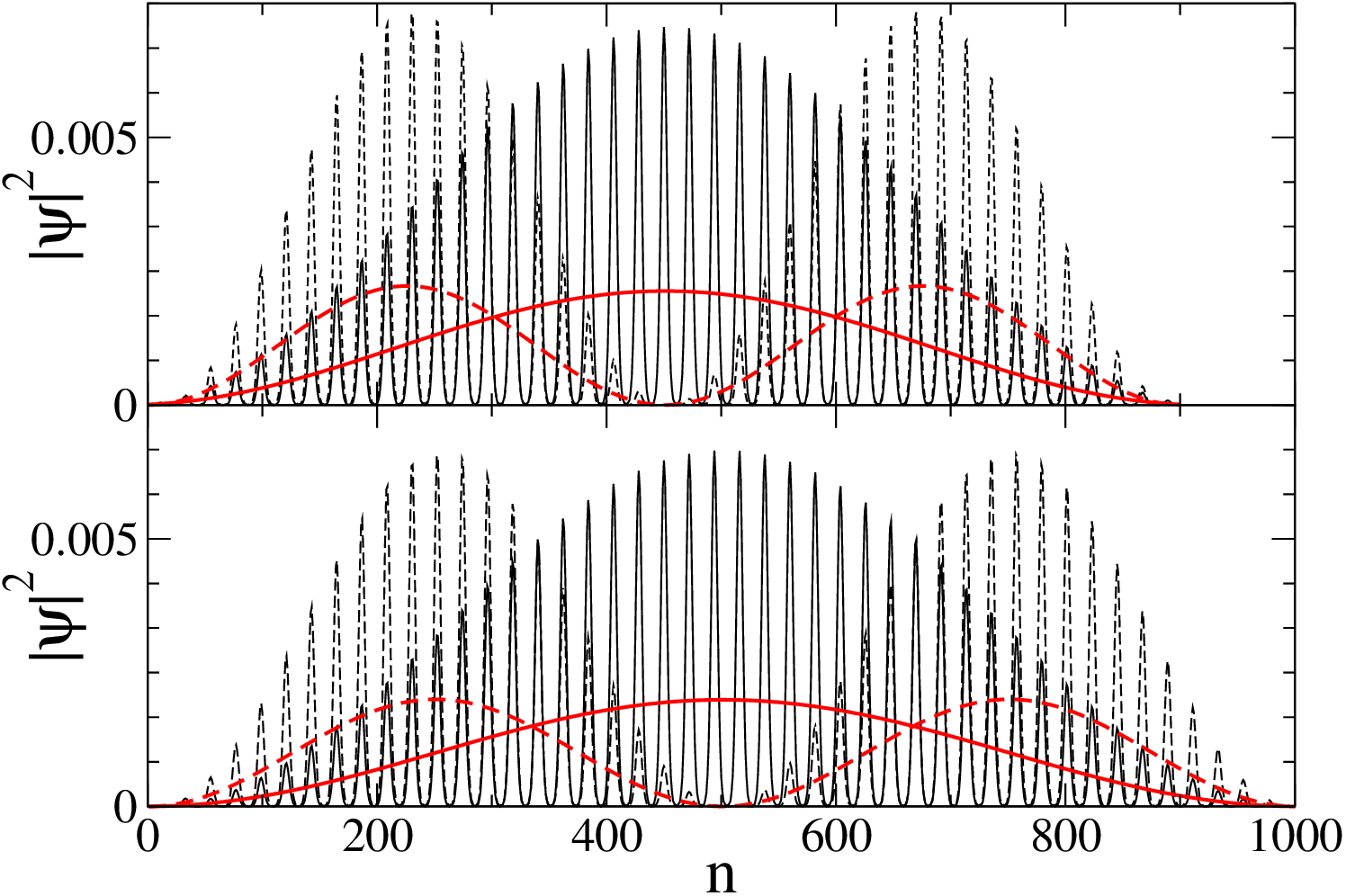}
\vskip 0.2truecm
\caption{\label{fig2}
(Color online)
The ground and first excited state wave functions 
for $L=900$ (upper panel) and $L=1000$ (lower panel).
For the Harper model the ground state wave function squared
corresponds to the black continuous line, while the dashed
curve corresponds to the first excited state. For the clean
ring with an impurity the same notation is used with red curves.
For the Harper model the Gaussian localized states at the
superlattice points are seen. It is also evident that
the Harper states have a similar envelope to the clean state
with an impurity.
}
\end{figure}

Thus, to first approximation, there is a superlattice 
of valleys at points $j_n$, with $n=1,\ldots, L_n$, each with a 
zero-energy state, $|n\rangle$, 
centered around $j=j_n$. These states 
can be written as Gaussians falling off at a length scale of
$\xi=\sqrt{t/\pi\lambda|\epsilon|}$. 
The central band eigenfunctions are composed of
the hybridization of these localized state.
For periodic boundary conditions, and when $L_n = \lfloor L_n \rfloor$, 
the eigenstates of the central band 
are plane waves composed of the valley Gaussian,
$|k\rangle = L_n^{-1/2} \sum_{n=1}^{L_n} S_n e^{ikn} |n\rangle$,
where 
$S_n = \sqrt{2} \cos \left(n\pi/2-\pi/4\right)=\ldots,1, 1,-1,-1, 1, 1,\ldots$.
The spectrum $E^{\rm central}(k) = -2\bar{t}\cos k$ \cite{kraus14}, 
where $k = 2\pi m/L_n$ ($m=0,\pm 1,\pm 2, \ldots$)
and $\bar t \approx \exp(-\xi^2/(4\xi^2\epsilon)^2)\big(2t 
\exp(-1/4\xi^2)\sinh[(4\xi^2|\epsilon|)^{-1}] - 
\lambda \exp(-\pi^2 \epsilon^2 \xi^2)\big)$. 
Thus,
the central band spectrum is expected to show a degeneracy since
$E^{\rm central}(k)= E^{\rm central}(-k)$.
A closer look at this issue reveals that if the
system is not {\it exactly} periodic, the degeneracy
will be broken by the non perfect periodicity. 
One may think of the effect of the non perfect periodicity
as an impurity at the region of the non-periodicity 
(i.e., around $n=0$). For low-lying states in the central
band, the impurity acts as a hard-wall, leading to
low-lying states of the form:
$|\tilde {k}\rangle = \sqrt{2/L}\sin(\tilde {k} n) $,
with $\tilde {k} = \pi m/L_n$ ($m=0,1,2,\ldots$), and
eigenvalues $E^{\rm central}(\tilde k) = -2\bar{t}\cos \tilde k$.
Thus, as can be seen in the inset  of Fig. \ref{fig0},
for the low-lying states in the central
band the degeneracy in the eigen-values is lost,
both for the almost periodic case ($L=900$), as
for the non-periodic one ($L=1000$). The
low-lying wave functions are depicted in the
upper panel of Fig. \ref{fig2}. For comparison
the wave functions of a clean ring of the same
length, with a single impurity at $n=0$
(weak for $L=900$, strong for $L=1000$), is drawn.
For the clean system the
ground state wave function corresponds to a half-sine,
while the first excited state to a sine. This behavior
is not very sensitive to the impurity strength.
A similar situation can be seen for the Harper model,
where the half-sine and sine envelopes are composed
of the Gaussian superlattice states at points $j_n$.
As expected for the low-lying states there is no essential
difference between the almost periodic case and the
non-periodic one.

\begin{figure}
\includegraphics[width=8.5cm]{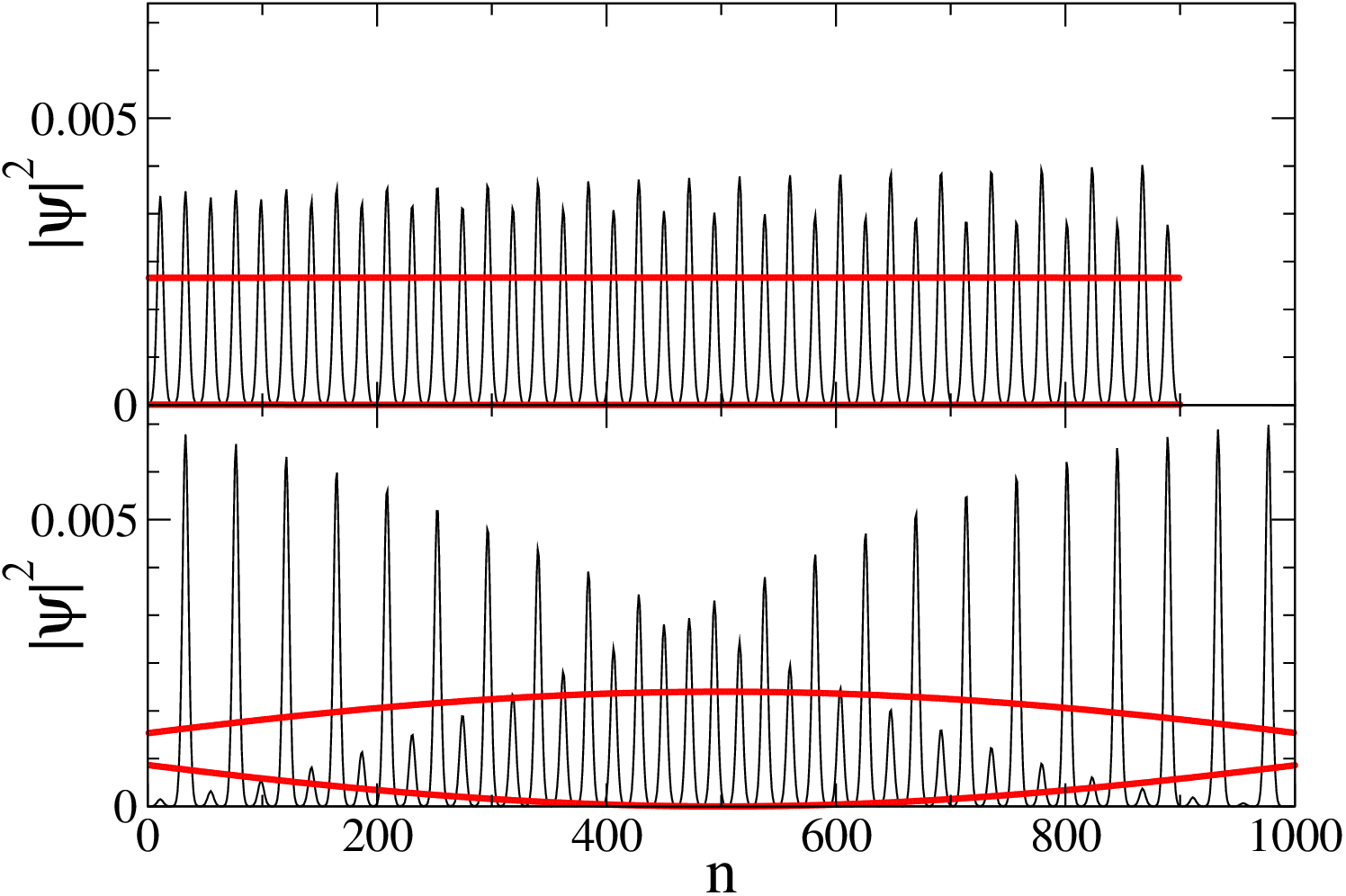}
\vskip 0.2truecm
\caption{\label{fig2a}
(Color online)
The $L/2+1$ excited state wave function for the Harper (black curve) and
clean system with an impurity (red symbols), for $L=900$ (upper panel) 
and $L=1000$ (lower panel). For $L=900$ which is almost commensurate 
(and the corresponding clean system with weak impurity) the wave function
is more or less homogeneous. For $L=1000$ which is not commensurate 
(and the corresponding clean system with a strong impurity) the wave function
shows a spatial structure. In both cases the clean system also shows a fast
$(-1)^n$ modulation since it is at half filling.
}
\end{figure}

This changes when higher levels are considered.
For higher energies, one expects that the effect
of the impurity will be weaker. Indeed, as can be
seen for the clean system depicted in the lower panel 
of Fig. \ref{fig2a}, if the strength of the impurity is
weak (the $L=900$ case), the wave function is homogeneous.
Only for the stronger impurity (the $L=1000$ case), the wave function 
shows a signature of the impurity. A similar behavior
is seen for the central band of the Harper model.
For the almost commensurate case the states in the
middle of the central band show only a slight degeneracy 
breaking (see the inset of Fig. \ref{fig0}, where the energies
for $L=900$ clearly appear in pairs). Nevertheless,
when the non-commensurability is stronger, as for
$L=1000$, the levels are non-degenerate even in the
middle, thus corresponding to a hard wall boundary condition
even for higher energies. This distinction
can be seen also for the wave functions, where for the almost 
commensurate Harper model the wave function is almost homogeneous,
while for the non-commensurate case the wave function is non-homogeneous
(see Fig. \ref{fig2a}).

\begin{figure}
\includegraphics[width=8.5cm]{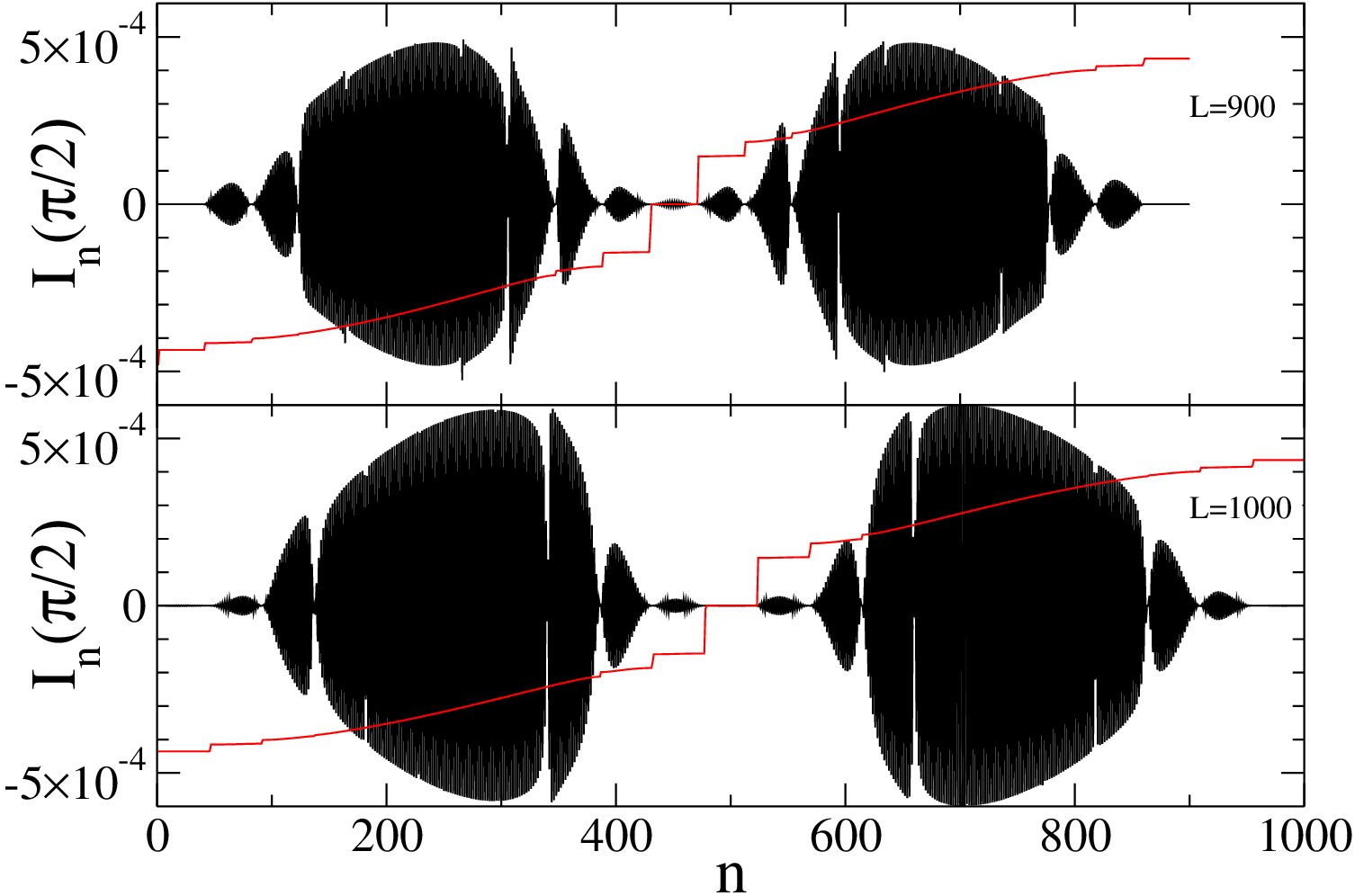}
\vskip 1.4truecm
\includegraphics[width=8.5cm]{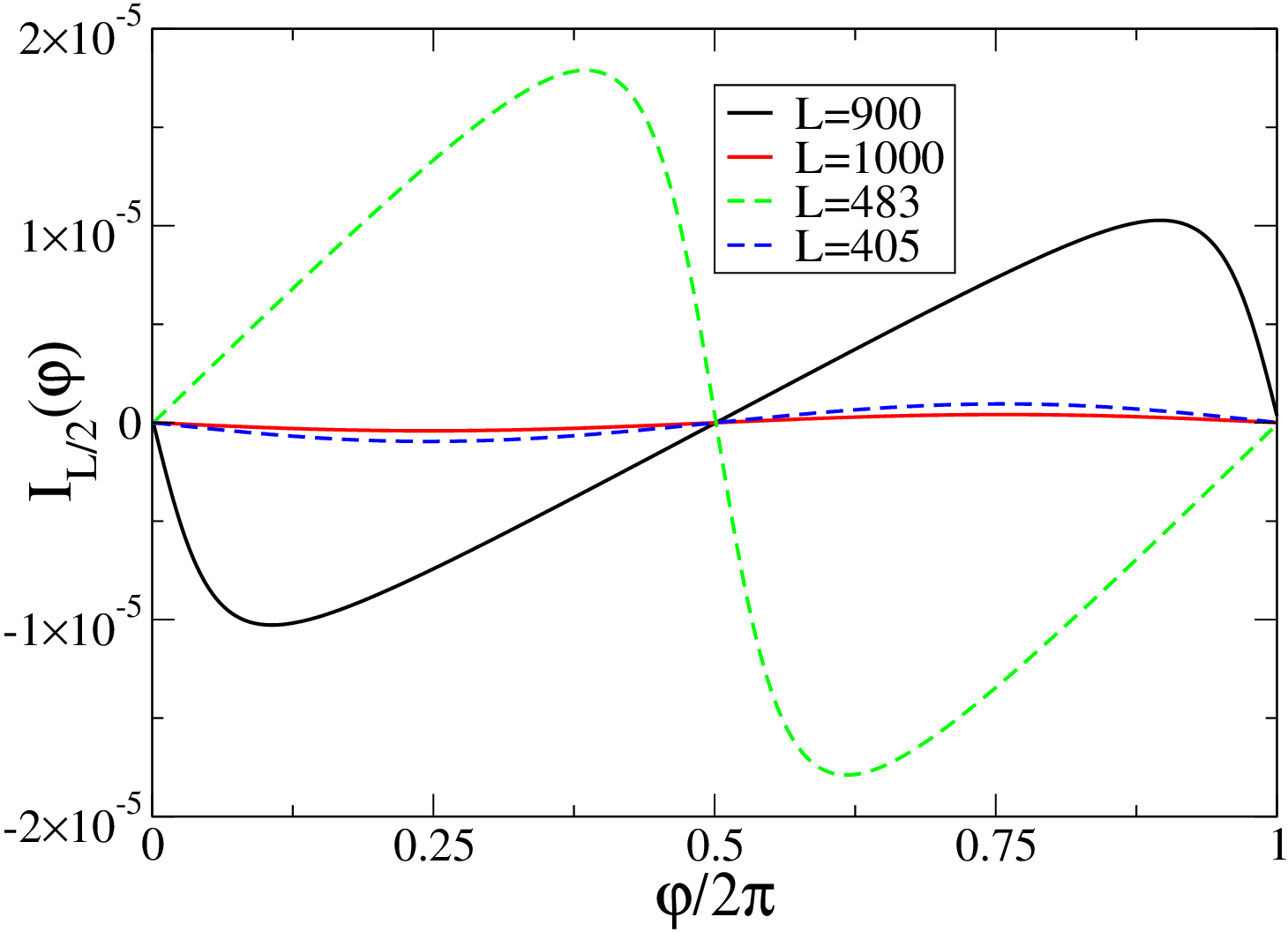}
\caption{\label{fig5}
(Color online)
The total persistent current 
for b=$\sqrt30$ ($\epsilon=-0.22774$) 
and $\lambda=1$ for different sample length.
Upper panel:
The total PC at a given flux $I_n(\phi=\pi/2)$ as function of the
number of states $n$ for the $L=900$ and $L=1000$ length is represented
by the black lines. The red curve corresponds (with an arbitrary scale)
to the energy of the $n$-th level. Jumps in the red curve indicates
a gap between the states. Close to the gaps (i.e., at the band edge) the
total PC goes to zero. The main difference between the almost
commensurate ($L=900$) and non commensurate ($L=1000$) length
is the amplitude of the total PC in the central band (the states
around $L/2$).
Lower panel:
The total PC at the center of the central band as function of the threading
flux $\phi$. 
For the almost commensurate superlattice cases ($L=900$, and $L=483$)
the PC is large and it has a saw tooth like shape. For the
incommensurate cases ($L=1000$, and $L=405$) the PC is small
and the shape is sine like.
}
\end{figure}

Let us now consider the effects of commensurabilty 
on the persistent current following through a Harper ring. 
The PC of the $j$-th level is defined as
$\iota_j(\phi)=\partial \varepsilon_j / \partial \phi|_{\phi}$, while
the total persistent current up to a given level $n$
is $I_{n}(\phi)=\sum_{j=1}^{n} \iota_j(\phi)$.
The total persistent current for any value of $n$
for length $L=900$ and $L=1000$
are presented in Fig. \ref{fig5}, with the data pertaining
to the energy of the $n$-th level. Gaps between 
the bands are clearly seen by the jumps in the 
energy between adjacent levels (as in Fig. \ref{fig0}). 
First one notes
that as expected, the total persistent current
reaches a maximum at the middle of the bands, while
it is zero at the edge of the band.
For the bands far from middle, there is no essential difference 
in $I_n$ between the almost commensurate ($L=900$)
length and the incommensurate length ($L=1000$).

Since we are considering here the metallic regime
of the Harper model ($\lambda=1$), a metallic behavior
of the PC is expected. The most obvious difference
between the PC of a metallic and a localized system
is the amplitude of the PC. As can be seen from 
the energy dependence on the threading flux presented
in Fig. \ref{fig5}, there is a huge difference 
in the dependence of the states between
almost commensurate length ($L=900$ and $L=483$)
and incommensurate length ($L=1000$ and $L=405$).
In this figure the flux dependence of 
total PC $I_{L/2}(\phi)$ is plotted.
For the commensurate case a high amplitude
of the PC is seen, which is in line with our expectations
from a metallic system. Moreover,
a unique sawtooth dependence
as function of the flux is observed, similar to the typical
behavior of the total PC in clean and weakly disordered 
systems \cite{cheung88}. 
On the other hand,  for the incommensurate case $I_{L/2}(\phi)$ is 
almost flat and has a sine flux dependence, expected in the
localized regime. 

\begin{figure}
\includegraphics[width=8.5cm]{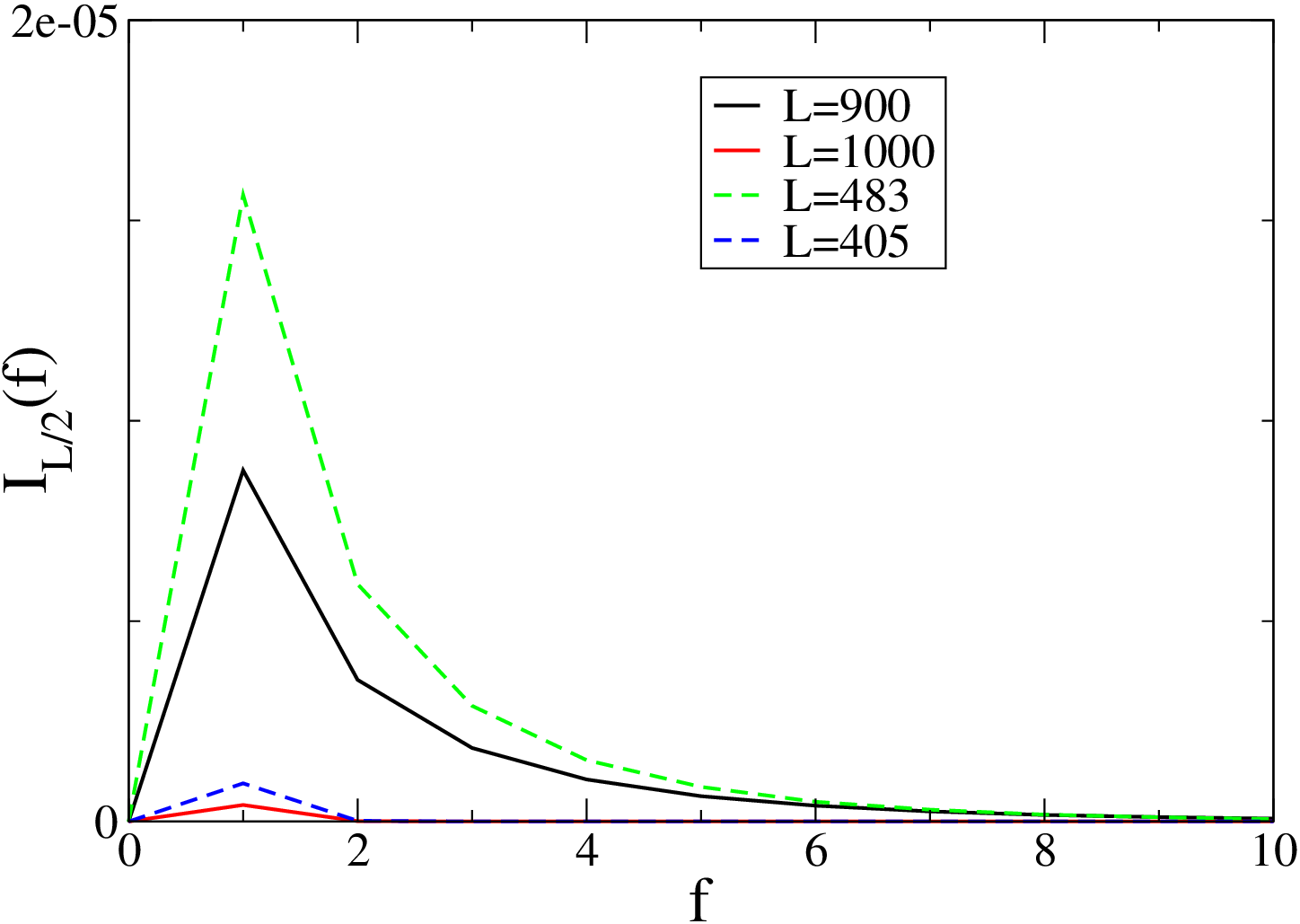}
\caption{\label{fig6}
(Color online)
Fourier transform of the total PC at the middle 
of the spectrum. 
For the
incommensurate cases ($L=1000$, and $L=405$) the Fourier
transform has only one significant component which is relatively small.
For the almost commensurate superlattice cases ($L=900$, and $L=483$)
the first component is the biggest, but the contribution from the
higher components goes down only gradually.
}
\end{figure}

An additional indicative feature for determining whether 
the PC corresponds to a
metallic or a localized behavior, is the number
of harmonics composing the PC. Generally, the PC 
may be written as a harmonic expansion of the
form $\iota=\sum_f \iota(f) \sin(2 f \pi \varphi/\varphi_0)$
and $I=\sum_f I(f) \sin(2 f \pi \varphi/\varphi_0)$. 
For the localized 
regime the PC is described by the first harmonic, i.e.,
$\iota \sim \iota(1) \sin(2 \pi \varphi/\varphi_0)$ and
$I \sim I(1) \sin(2 \pi \varphi/\varphi_0)$ \cite{cheung88},
while for the metallic regime, the weight of higher
harmonics falls off rather slowly \cite{bouchiat89}. Thus, 
both the amplitude and dependence on the flux of the
almost commensurate and non-commensurate length demonstrate
a completely different behavior at the middle of the central
band. This is also evident from the Fourier tranform of the
PC (Fig. (\ref{fig6}), where for the non-commensurate cases
only the first harmonic is non-zero, while for 
almost commensurate cases the contribution of higher
harmonics is still significant.

\begin{figure}
\vskip 1truecm
\includegraphics[width=8.5cm]{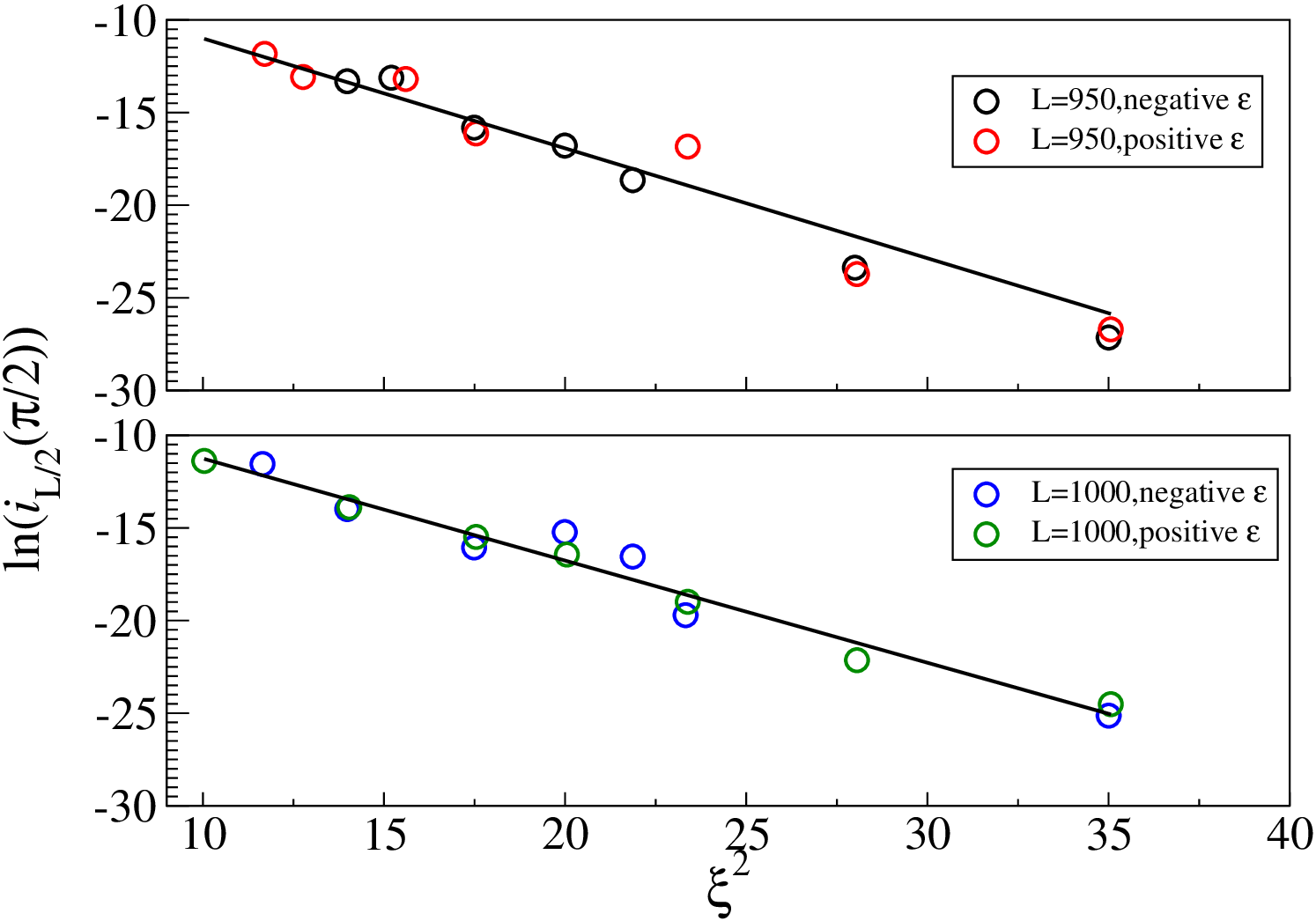}
\vskip 1truecm
\caption{\label{fig4}
(Color online)
The first harmonic of the central state PC $\iota_{L/2}(1)$ as a 
function of $\xi^2$,
with b=$5.5+\epsilon$ and $\lambda=1$. In order to change the width
of the Gaussian states on the superlattice points ($\xi^2$) we change
$\epsilon$ by either positive or negative values. It is clear that
$\iota_{L/2}(1)$ depends on $|\epsilon|$, and is well described by an
exponential decrease with $\xi^2$.
upper panel: $L=950$
Lower panel: $L=1000$} 
\end{figure}

Therefore, if one uses the total PC at half-filling, or
the persistent current of a particular state in the middle
of the central band to
decide whether the Harper model is metallic or localized,
one would get different results as function of the length of
the sample. The behavior is metallic for the almost commensurate length and
localized like for the non commensurate length. Of course the
system remains essentially metallic in both cases.
Nevertheless, as we have seen from a direct inspection of
the wave functions in the central band (Figs. \ref{fig2},\ref{fig2a})
the non commensurate segment has an influence similar to an
impurity embedded in the ring. Since this impurity
acts like a barrier, we expect that the PC amplitude
will be suppressed proportionally to the overlap between
two superlattice states, i.e., proportionally to $\exp(C/\xi^2)$
(where $C$ is a constant). 
In Fig.\ref{fig4} we plot
the first harmonic of the PC of the middle state $\iota_{L/2}(1)$
is plotted as function of $\xi^2=t/\pi\lambda|\epsilon|$. 
Here we keep the strength of the potential $\lambda$ and
system size $L$ constant, while 
changing $|\epsilon|$. It can be seen that 
$\iota_{L/2}$ is suppressed exponentially as function of $\xi^2$.
This fits well with the description of the non-commensurate central
band as a system with an impurity. It also confirms that 
PC can not be used in these systems to determine whether the
system is metallic or localized.

In conclusion, we have considered the PC in the 
staggered Harper model, mainly close to the half-filling.
Contrary to naive expectations the PC for most systems
do not show typical metallic behavior, although the system
is in the metallic regime. This stems from the unique character
of the central band states, which are a hybridization 
of Gaussian states localized in super lattice points. If the
distance between the superlattice sites are (almost) commensurate
with the system size, then the PC has metallic properties. On the
other hand, when the superlattice sites are incommensurate
the system has an effective impurity at the incommensurate section and
the PC shows the signature of an insulator. Since for most
combinations of the Harper frequency $b$ (close to the staggered
condition $\mathbb{Z}+1/2$) and length $L$ are incommensurate,
a typical Harper model will actually exhibit PC corresponding to
an insulator.

Thus, studying the PC at half-filling for the finite almost staggered
Harper model is not indicative of the phase of the system. 
Moreover, also other stout method for identifying the metal-insulator
transition, such as the level spacing statistics \cite{shklovskii93, berkovits96a},
will not show the expected transition in the level statistics
from the Poisson distribution for the localized regime to the
Wigner distribution in the metallic regime. As is evident from the
inset in Fig. \ref{fig0}, the distribution will change even in the
metallic regime, between a picket fence distribution to a bi-model one.
Therefore, when studying the Harper model in order to investigate
the metal-insulator transition one must be careful to choose
the appropriate measures.


\subsection*{Acknowledgments}
Financial support from the Israel Science Foundation (Grant 686/10) is
gratefully acknowledged.
%

\end{document}